\documentclass[doublecol]{epl2}

\usepackage{amsmath}
\usepackage{amssymb}
\usepackage{graphicx}
\usepackage{calrsfs}

\newcommand{\be}{\begin{equation}}
\newcommand{\ee}{\end{equation}}
\newcommand{\ben}{\begin{equation*}}
\newcommand{\een}{\end{equation*}}

\newcommand{\Li}[1]{\mathrm{Li}_{#1}}

\newcommand{\re}{\Re\mathrm{e}}
\newcommand{\im}{\Im\mathrm{m}}
\newcommand{\dd}{\upd}
\newcommand{\rmTE}{\mathrm{TE}}
\newcommand{\rmTM}{\mathrm{TM}}

\newcommand{\kB}{k_\mathrm{B}}
\newcommand{\rs}{r_\sigma}
\newcommand{\us}{u_\sigma}

\newcommand{\infintw}{\int_{0}^{\infty}\!\!\! \dd \omega}
\newcommand{\ssum}{\sum_{\sigma=\rmTE}^\rmTM}

\title{Frequency spectrum of the Casimir force: interpretation and a paradox}
\shorttitle{Frequency spectrum of Casimir forces}
\date{\today}
\author{S. A. Ellingsen }
\institute{Department of Energy and Process Engineering, Norwegian University of Science and Technology, N-7491 Trondheim, Norway}
\pacs{34.20.-b}{Interatomic and intermolecular potentials and forces}
\pacs{12.20.-m}{Quantum electrodynamics}
\pacs{03.70.+k}{Theory of quantized fields}

\abstract{The frequency spectrum of the Casimir force between two plates separated by vacuum as it appears in the Lifshitz formalism is reexamined and generalised as compared to previous works to allow for imperfectly reflecting plates. As previously reported by Ford [Phys. Rev. A \textbf{48} (1993) 2962], the highly oscillatory nature of the frequency dependence of the Casimir force points to possibilities for very large and indeed negative Casimir forces if the frequency-dependent dielectric response, $\epsilon(\omega)$, of the materials could be tuned. A paradox occurs, however, because an alternative calculation of the effect of a perturbation of $\epsilon(\omega)$ involving a Wick rotation to imaginary frequencies indicate only very modest effects. A recent experiment appears to convincingly rule out the reality of Ford's optimistic predictions, although given the enormous technological promise of such frequency effects, further theoretical and experimental study is called for.}

\begin{document}

\maketitle

In an interesting paper \cite{ford93}, Ford analysed the frequency spectrum of the Casimir pressure as it appears when read directly out of Lifshitz' celebrated formula\cite{lifshitz56}. His calculations extended a previous study of the spectrum of the Casimir effect for a massless scalar field \cite{ford88} and subsequent analysis of the electromagnetic vacuum stress tensor by Hacyan et al.~\cite{hacyan90}. In \cite{ford93}, the classical Casimir set-up is considered, where two perfectly reflecting metal plates of infinite transverse size are separated by a vacuum-filled gap of width $a$. For this system, the pressure between the plates was found by Casimir \cite{casimir48} to be
\be\label{Casimir}
  P_C(a) = -\frac{\hbar c\pi^2}{240a^4}.
\ee
Ford's puzzling finding was that if the pressure is expressed as an integral over all frequencies of the zero-point oscillations of the electromagnetic field in the cavity, the integrand is wildly oscillating and discontinuous as a function of frequency and the integral a sum of almost exactly cancelling positive and negative contributions, each of which far larger in magnitude than the measurable pressure itself. By a suitable cutoff procedure, however, the integral is calculable and the result correct. Similar considerations were later performed for a sphere and plate set-up \cite{ford98, sopova04} and the electromagnetic stress tensor in a cavity\cite{ford07}. An extension of Ford's work on two ideally conducting plates was recently presented by Lang \cite{lang05}.

The unruly behaviour of the Casimir force as a function of real frequencies was recently treated for numerical purposes\cite{rodriguez07} and the same oscillatory behaviour was found. As a consequence these latter authors like most before performed the Wick rotation to imaginary time (and imaginary frequencies), which is legal when the permittivity is assumed causal. As expressed for imaginary frequencies, the Lifshitz expression is much more well behaved and rather than complex and strongly oscillating, the frequency integrand is real, nicely peaked and exponentially decreasing at high imaginary frequency. The importance of good optical data for the precise calculation of Casimir forces has been emphasised in a number of recent efforts \cite{lambrecht00,pirozhenko06,pirozhenko08, svetovoy08}, but these have all employed a Wick rotated formalism.

Ford suggested that if the frequency response of the plate materials could be tuned, for example if a material could be found which is transparent for all but a small band of frequencies in which it was a good reflector, Casimir forces much larger than that between perfect conductors could be observed and by changing the reflection band the force could be changed from attractive to repulsive. Despite the potentially enormous technological potential of such tuning for applications in microengineering and nanotechnology, the issue of the physical interpretation of the frequency spectrum of the Casimir force has remained largely unaddressed.

A relevant experiment was recently performed by Iannuzzi, Lisanti and Capasso at Harvard\cite{iannuzzi04} in which the Casimir force was measured in the same configuration with a good and poor reflector respectively. The material used was a so-called hydrogen-switchable mirror which can be switched from mirror to transparent at optical frequencies by introducing hydrogen. At frequencies in the IR and UV parts of the spectrum, reflection is presumed by the authors of \cite{iannuzzi04} to be approximately unchanged. This is the inverse of the situation suggested by Ford, and based on \cite{ford93} large effects should be expected. Iannuzzi et al.\ observed no change of the force in the two cases, however. 

The paradox is theoretical as well as experimental: Assume that the permittivity of one or both of the plates in a standard two-plate set-up \cite{casimir48} is changed in a band of frequencies. The effect on the force of this perturbation may, one may think, be calculated in two different ways. Either Lifshitz' expression for the force is integrated over the relevant band of \emph{real} frequencies and the difference taken. Alternatively, the new, perturbed permittivity is rotated to imaginary frequencies using the Kramers-Kronig relations and the force calculated as an integral (assuming zero temperature) over imaginary Matsubara frequencies. The numerical results obtained in the two ways, however, appear to differ greatly.

In the following, Ford's calculations are generalised to imperfectly reflecting plates represented by an effective reflection coefficient. While the same oscillatory behaviour is found, the integrand is no longer discontinuous. In the limit of perfect reflection, Casimir's result is once more obtained. Thereafter the paradox is elaborated and compared to the recent experiment by Iannuzzi et al. 

%%%%%%%%%%%%%%%%%%%%%%%%%%%%%%%%%%%%%%%%%%%%%
\section{Casimir force and frequency spectrum with constant reflection coefficients}

\begin{figure}[htb]
  \centerline{\includegraphics[width=1.2in]{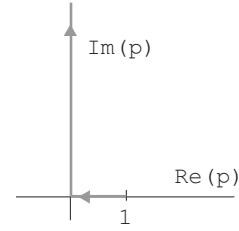}}
  \caption{The integration contour in (\ref{Lifshitz}).}
  \label{fig_contour}
\end{figure}

The expression for the Casimir pressure was given by Lifshitz as an integral over (real) frequencies and the variable $p$ by\cite{lifshitz56}

\begin{align}\label{Lifshitz}
  P_T(a) =& -\frac{\hbar}{2\pi^2 c^3}\re \int_0^\infty \dd \omega\omega^3 \int_C \dd p p^2 \coth\left(\frac{\hbar \omega}{2\kB T}\right)\notag\\
  &\times\sum_{\sigma=\rmTE}^\rmTM \frac{r^2_\sigma \exp(2ip\omega a/c)}{1-r^2_\sigma \exp(2ip\omega a/c)}
\end{align}
where the variable $p$ relates to the transverse wave-vector $\mathbf{k}_\perp$ as 
\ben
  p=\frac{i c}{\omega}\sqrt{\mathbf{k}_\perp^2-\omega^2/c^2}
\een
and the contour of integration $C$,  shown in figure \ref{fig_contour}, runs from 1 to 0 along the real axis and thence along the imaginary axis to $+i\infty$. We have assumed both plates equal in (\ref{Lifshitz}) but all results in the following may be generalised to different media 1 and 2 by letting $r^2\to r_1r_2$. It is easy to show that the former part of the integral covers all modes which propagate in vacuum, that is when $\omega > |\mathbf{k}_\perp|c$, and the latter covers all evanescent modes $\omega < |\mathbf{k}_\perp|c$ which vanish exponentially away from the surfaces. 

In the first part of the analysis we shall consider complex reflection coefficients which are constant with respect to $\omega$ and $p$, but in principle dependent on the separation $a$. 
We will assume zero temperature throughout so that the coth function in (\ref{Lifshitz}) is set to 1.

To denote the frequency dependence we use the notation
\ben
  P = \infintw P_\omega = \infintw \ssum P_{\omega\sigma}.
\een
Let us assume now that the reflection coefficient $r_\sigma$ is a constant quantity with respect to $\omega$ and $p$ which we allow to be complex for generality.

\begin{figure}[t]
  \centerline{\includegraphics[width=2.5in]{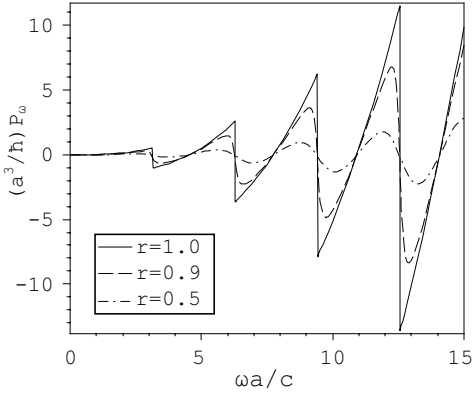}}
  \caption{Plot of (\ref{Pws}) normalised by $a^3/\hbar$ as a function of $\omega a/c$ for different real values of $r$. This figure generalises figure 2 of \cite{ford93}.}
  \label{fig_graph}
\end{figure}

Consider first the propagating part of (\ref{Lifshitz}). The $p$ integral may now be evaluated explicitly. Consider one of the modes and suppress the index $\sigma$ for now to find
\begin{align*}
  \re\int_1^0 &\frac{\dd p\cdot p^2 r^2 e^{2 i p\omega a/c}}{1-r^2 e^{2 i p\omega a/c}}=\re\sum_{n=1}^\infty \frac{\partial^2}{\partial \gamma_n^2}\frac{r^{2n}}{ i\gamma_n}( e^{ i\gamma_n}-1)\\
  &= \re\sum_{n=1}^\infty r^{2n}\left[\frac{ i e^{ i n\xi}}{n\xi}-\frac{2 e^{ i n\xi}}{n^2\xi^2}-\frac{2 i}{n^3\xi^3}( e^{ i n\xi}-1)\right]
\end{align*}
where $\gamma = n\xi$ and $\xi = 2\omega a/c$. We now introduce the polylogarithmic function whose $m$th order is defined
\ben
  \Li{m}(z) = \sum_{n=1}^\infty \frac{z^n}{n^m}.
\een
Then we may write the pressure contribution from propagating waves (PW) $P_{\omega\sigma}^\text{PW}$ as
\begin{align}
  P^\text{PW}_{\omega\sigma}=&\frac{-\hbar}{16\pi^2a^3}\left[-\xi^2\im\Li{1}(\rs^2 e^{ i\xi})-2\xi\re\Li{2}(\rs^2 e^{ i\xi})\right.\notag\\
  &\left. +2\im\Li{3}(\rs^2 e^{ i\xi})-2\im\Li{3}(\rs^2) \right].\label{Pws}
\end{align}
When $\rs$ are real, there is no evanescent contribution to the pressure as we will see. We have plotted the integrand (\ref{Pws}) as a function of $\xi$ for various real values of the effective reflection coefficient $r$ in figure \ref{fig_graph} (assumed equal for TE and TM for simplicity). In the limit $r=1$ one obtains the discontinuous behaviour reported by Ford. Equation (\ref{Pws}) thus generalises Ford's calculation; indeed figure \ref{fig_graph} is formatted so as to ease comparison with figure 2 of \cite{ford93}.

We can go on to calculate the force for constant reflection coefficients. The analysis so far only required $\rs$ to be independent of $p$; one could define some $p$ averaged $r_\sigma(\omega)$ and plot it the way we have in figure \ref{fig_graph}. By assuming $r$ independent of $\omega$ as well, however, we are able to evaluate the $\omega$ integral explicitly. We substitue $\omega$ by $\xi$ in the integral and write $u_\sigma\equiv \rs^2\exp( i\xi)$ for short.

For a $\xi$ independent $\rs$ one may show from the recursion property\cite{kolbig70}
\ben
  \int_0^z \dd t \frac{\Li{n}(t)}{t} = \Li{n+1}(z)
\een
that the following relations are fulfilled:
\begin{align*}
  \im\int_0^\xi \dd\xi' \Li{2n-1}(\rs^2 e^{ i\xi'})&=-\re\Li{2n}(\rs^2 e^{ i\xi})\\
  \re\int_0^\xi \dd\xi' \Li{2n}(\rs^2 e^{ i\xi'})&=\im\Li{2n+1}(\rs^2 e^{ i\xi}).
\end{align*}
The integrand as given by (\ref{Pws}) is a wildly oscillating function of $\xi$ as figure \ref{fig_graph} indicates. For definiteness, let $r\to r\exp(-\delta\xi)$ where $\delta$ is a small real quantity which we will take to zero in the end. The same procedure was followed by Ford \cite{ford93} and reflects the physical fact that all materials become transparent at very high frequencies. By moderately lengthy but straightforward partial integrations of (\ref{Pws}) with respect to $\xi$ one may then obtain for constant and real reflection coefficients
\begin{align}
  P^\text{PW}_\sigma =& \frac{-\hbar c}{16\pi^2a^4}\left[\frac{1}{2}\xi^2\re\Li{2}\us - 2\xi\im\Li{3}\us\right.\notag\\
  &\left.-3\re\Li{4}\us -\frac{1}{2}\xi\im\Li{3}\rs^2\right]_{\xi=0}^\infty.\label{Pint}
\end{align}
With the normalising factor $\exp(-\delta\xi)$ in all reflection coefficients the $\infty$ limit is zero as physically expected. In the limit $\xi=0$, $\us\to \rs^2$ and only the term with no factor $\xi$ remains:
\ben
  P^\text{PW}_\sigma = -\frac{3\hbar c}{16\pi^2 a^4}\re \Li{4}\rs^2.
\een

Secondly we consider the part of the pressure from evanescent waves (EW). The same way as above we evaluate the $p$ integral thus:
\begin{align}
  \re\int_0^{i\infty}\frac{\dd p\cdot p^2}{\rs^{-2} e^{-2 i pwa/c}-1} &=  \im\int_0^{\infty}\frac{\dd q\cdot q^2}{\rs^{-2} e^{2qwa/c}-1}\notag\\
  &= \frac{2}{\xi^3}\im\Li{3}(\rs^2)\label{evanint}
\end{align}
where $p= i q$. Clearly if $\rs$ is real as in a non-dissipative medium, there is no evanescent contribution. Moreover, one notices that that the evanescent contribution is exactly cancelled by the last term of (\ref{Pws}) above. Using the same normalisation as before the evanescent part accumulates to zero. Note that we could have let the normalisation $\exp(-\delta\xi)$ pertain to the exponential factor $\exp( i \xi)$ in the integration (\ref{Pint}) since the remaining divergent term is exactly cancelled by evanescent contributions.

Hence the final result is obtained when $\rs$ is constant:
\begin{equation}\label{rCasimir}
  P(a,r) = -\frac{3\hbar c}{16\pi^2a^4}\ssum \re \Li{4}(\rs^2).
\ee
In the limit of perfect reflection, $\rs\to 1$, the summation over $\sigma$ gives a factor 2, and with $\Li{4}(1) = \zeta(4) = \pi^4/90$ we get Casimir's result (\ref{Casimir}). $P(a,r)$ is plotted in figure \ref{fig_Pr} relative to the case for ideal reflection.
\begin{figure}[t]
  \centerline{\includegraphics[width=2.2in]{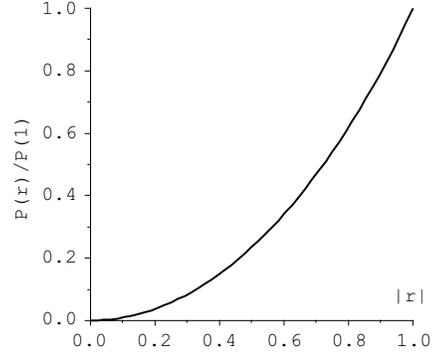}}
  \caption{The Casimir pressure for non-unitary real reflection coefficient (assumed equal for TE and TM) plotted relative to Casimir's result (\ref{Casimir}).}
  \label{fig_Pr}
\end{figure}

In exactly the same manner we could find the Casimir free energy for real and constant reflection coefficients to be given as
\be\label{rF}
  \mathcal{F}(a,r) = -\frac{\hbar c}{16\pi^2a^3}\ssum \re\Li{4}(\rs^2).
\ee
Again Casimir's result is obtained in the limit $\rs=1$. Some caution should be excerted here since function $\Li{4}(z)$ has a branch cut along the real axis from $z=1$ to $\infty$ across which its imaginary part is discontinuous. Its real part, however, is continuous everywhere\cite{kolbig70}. 

%%%%%%%%%%%%%%%%%%%%%%%%%%%%%%%%%%%%%%%%%%%%%
\section{Physical significance: na\"ive approach}

Physically, reflection coefficients are far from constant with respect to frequency and transverse momentum, but the above analysis using constant reflection coefficients can be thought of as an averaging process in which the constant $\rs$ is defined as the value which, when replacing the realistic $\rs(\omega,p)$ does not change the value of the integral (\ref{Lifshitz}). Such an interpretation requires $\rs$ to depend on $a$ as well.

Using the calculated Casimir pressure between gold plates where realistic data from \cite{palik95} are employed (the calculation was previously presented in \cite{brevik06}), we can calculate the effective reflection coefficients $\rs(a)$ by 
\be\label{effref}
  \rs = \left[\Li{4}^{-1}\left\{-\frac{16\pi^2a^4}{3\hbar c}P^\text{num}_\sigma(a)\right\}\right]^{1/2}
\ee
where $P_\text{num}$ are the calculated data, and $\Li{m}^{-1}$ is the inverse polylogarithm of order $m$. By use of the series reversion functionality of analytic software the inverse polylogarithm is simple to calculate numerically. We plot the effective reflection coefficient in figure \ref{fig_effref}.

\begin{figure}[t]
  \centerline{\includegraphics[width=2.8in]{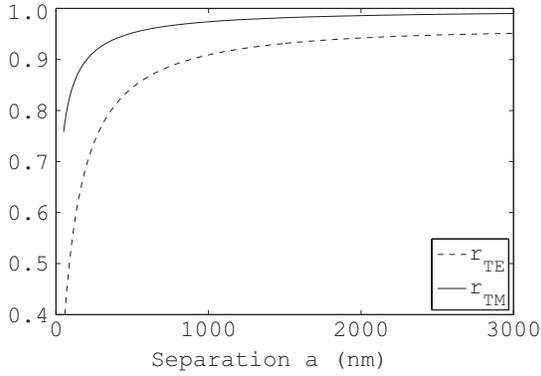}}
  \caption{$\rs(a)$ for gold as calculated from calculated force using (\ref{effref}).}
  \label{fig_effref}
\end{figure}

Ford \cite{ford93} (and later Lang \cite{lang05}) appears to argue that the frequency spectrum shown in figure \ref{fig_graph} resembles the real spectrum, with the modification that the oscillations will be dampened and finally vanish at high frequencies.  He concludes from this that since the oscillations themselves are much larger than the final force (which is the remainder of large fluctuations which cancel each other almost exactly), much larger and even repulsive Casimir forces could be obtained if media could be found which were good reflectors only in a range of frequencies.

\begin{figure}[t]
  \begin{center}
    \includegraphics[width=2.8in]{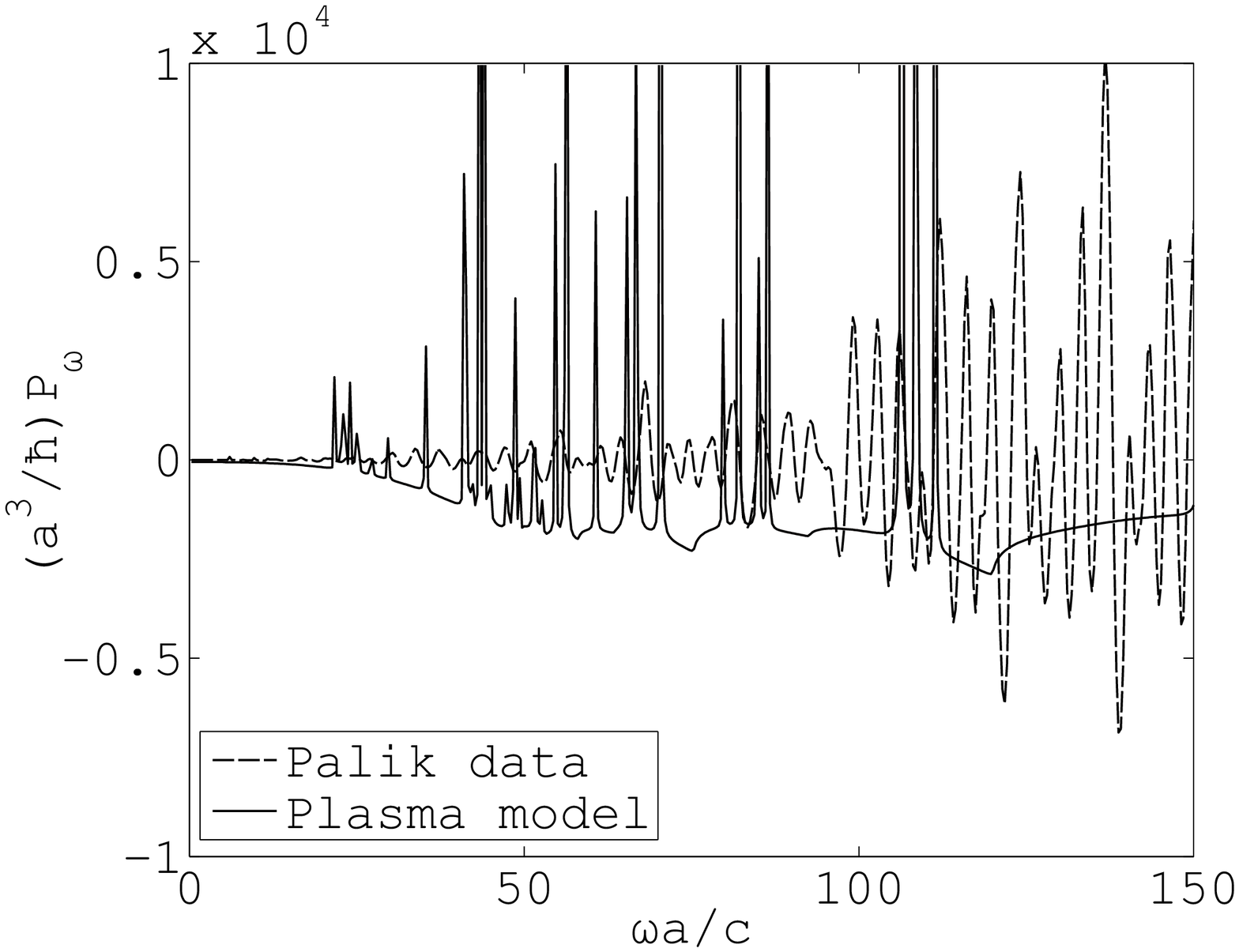}\\
    \includegraphics[width=2.8in]{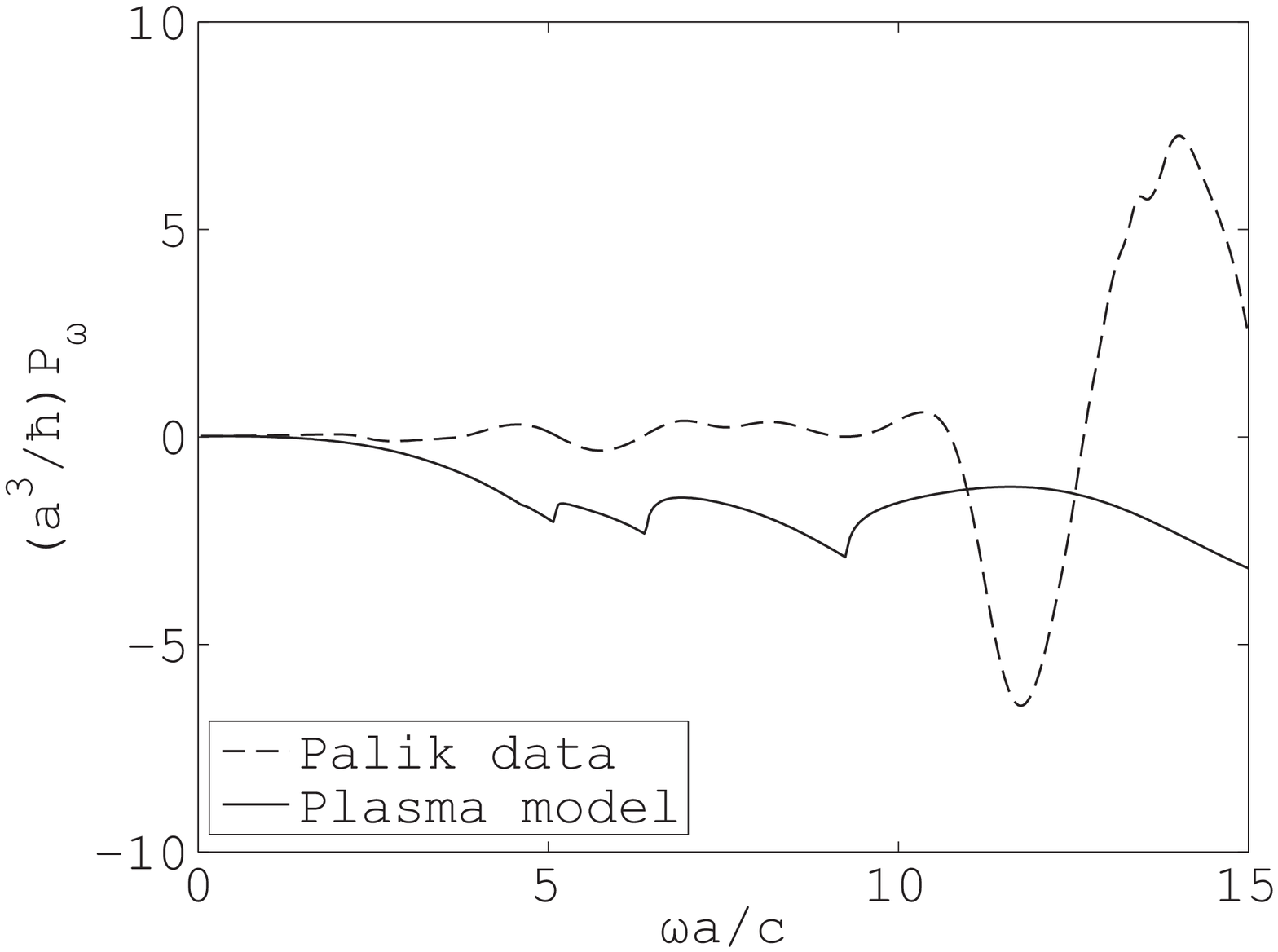}
    \caption{The frequency spectrum of the Casimir pressure calculated using real data for gold and the plasma model respectively. We have used $a=100$nm. The below graph is a zoom of the above.}
    \label{fig_palik}
  \end{center}
\end{figure}

Such a simple physical interpretation of the frequency spectrum is not unproblematic however. For comparison with a more physically realistic system we have plotted the frequency spectrum obtained by plugging tabulated optical data for gold from Palik's book \cite{palik95} directly into (\ref{Lifshitz}) and integrating over transverse momentum numerically (the separation is 100nm; the integrand is now a function of $\omega$ and $a$ individually, not only their product), as well as the same integrand as obtained when using the simple plasma model $\epsilon(\omega)=1-\omega_p^2/\omega^2$ using plasma frequency for gold $\omega_p=9$eV. Although the force predicted by either model at zero temperature are found to be very similar in magnitude when calculated by way of Wick rotation\footnote{Disregarding the disputed behaviour of the TE mode near zero frequency \cite{brevik06}.}, the frequency spectrum clearly differs greatly.

While a useful generalisation, the formalism of mean reflection coefficients should be used with care since its substitution renders the integrand of (\ref{Pint}) void of physical meaning other than giving the correct value per definition after integration: $\omega$ simply becomes a dummy variable. 
If Ford's result can be seen as a limiting case of the same procedure, this would indicate that the spectrum depicted in figure \ref{fig_graph}, while interesting may not represent physics. 

Nonetheless, the key feature of the integrands of figures \ref{fig_graph} and \ref{fig_palik}, the presence of large oscillations, is a hallmark of all these graphs. These fluctuations should be physically observable as noted by several authors. As we will see, however, indications are that while experimental confirmation of the Casimir force as calculated using Wick rotation is plentiful (see \cite{milton04} for a review), calculation by straightforward integration over real frequencies seems at odds with a recent experiment. This is paradixical since the two methods are typically presented as equivalent.

\section{The paradox and an experiment}

Assume that the permittivity of two metallic planes in a Casimir cavity is perturbed in such a way that it is made transparent in a band of frequencies but is still a good reflector outside this band (with reflectivity dying off at high frequencies). The effect on the force from this perturbation may be calculated in two different ways and the results are different.

Exactly such a situation was probed experimentally by Iannuzzi et al.~\cite{iannuzzi04}. The force between two metallic plates (in reality a sphere and a plane) was measured, one of which a hydrogen-switchable mirror (HSM). A good mirror in its as-deposited state, the HSM becomes transparent in the visible region once in a hydrogen rich atmosphere.
According to the authors of \cite{iannuzzi04} the switching of the mirror corresponded roughly to setting the material reflectivity to zero over a wavelength range $0.2-2.5$ $\mu$m, corresponding to $\omega$ between about $\omega_1=7.5\cdot10^{14}$1/s and $\omega_2=9.4\cdot 10^{15}$1/s. At a precision of 10-15\% (measured  roughly from figure 4 in \cite{iannuzzi04}) the group were unable to detect any difference in the force with the mirror switched on and off respectively.

We will estimate the effect of the transparency window using two different methods (an idealised version of such a material was considered in \cite{lang05}). Assume first that the boundaries of the transparent window, between frequencies $\omega_1$ and $\omega_2$, are sharp so that for a complex permittivity $\epsilon = \epsilon'+i\epsilon''\equiv 1+\chi$,
\ben
  \chi(\omega) \to \chi(\omega)[1-\theta(\omega-\omega_1)\theta(\omega_2-\omega)]
\een
where $\theta$ is the unit step function. Using, as in \cite{iannuzzi04}, the Drude model with data for gold, $\epsilon(\omega)= 1-\omega_p^2/(\omega^2+i\omega\nu)$ with $\omega_p=9$eV and $\nu=35$meV, the change in the permittivity at imaginary frequencies is fond as $\epsilon(i\zeta)\to \epsilon(i\zeta)-\Delta \epsilon(i\zeta)$ by use of the Kramers Kronig relation
\begin{align*} 
  \Delta \epsilon(i\zeta) =& \frac{2}{\pi}\int_{\omega_1}^{\omega_2}\frac{\dd \omega \epsilon''(\omega)}{\omega^2+\zeta^2}\\
  =&\frac{\omega_p^2}{\zeta^2-\nu^2}\frac{2}{\pi}\left[\arctan\frac{\omega_2}{\nu}- \arctan\frac{\omega_2}{\nu}\right.\\
  &-\left.\frac{\nu}{\zeta}\left( \arctan\frac{\omega_2}{\zeta}- \arctan\frac{\omega_1}{\zeta}\right)\right]
\end{align*}
the use of which ensures that the perturbation obeys causality. $\Delta\epsilon(i\zeta)$ makes for a correction to $\epsilon(i\zeta)$ on the level of 1\%-4\% and a corresponding correction to the force which would be unobservable at the precision of the experiment \cite{iannuzzi04}. The latter authors use precisely this argument to explain their negative result.

Now let us calculate the correction by instead inserting a modified $\epsilon(\omega)$ into (\ref{Lifshitz}) where for a single interface
\ben
  r_\rmTE = \frac{p-\sqrt{p^2+\epsilon-1}}{p+\sqrt{p^2+\epsilon-1}}; ~~r_\rmTM = \frac{\epsilon p-\sqrt{p^2+\epsilon-1}}{\epsilon p+\sqrt{p^2+\epsilon-1}}
\een
and instead of the step functions we model the transparent window using a function which allows smooth boundaries. Let $\chi(\omega)\to \chi(\omega)\phi(\omega; \Delta,s)$ with
\be
  \varphi(\omega) = 1-\frac{\Delta}{\pi}\left[\arctan\frac{s(\omega-\omega_1)}{c/a}+\arctan\frac{s(\omega_2-\omega)}{c/a}\right].
\ee
Here the parameter $\Delta\in [0,1]$ is the relative reduction in the window and $s$ determines the sharpness of the edges with $s=\infty$ giving unit step behaviour. The function $\phi$ is plotted in figure \ref{fig_phi} for $\Delta=1$ and different values of $s$.

We now calculate the change in the force by simply inserting this modified $\epsilon(\omega)$ into (\ref{Lifshitz}) and integrate over $p$ and a sufficiently large frequency range, then taking the difference. We use the Drude and plasma model respectively with parameters for gold to model $\epsilon$ (the data of \cite{palik95} contain too few points to use without extensive extrapolation). In the experiment \cite{iannuzzi04} only one of the plates was gold, yet the calculations are so rough that the difference does not matter here. The fact that in this estimation both reflection coefficients vanish in the transparency window whilst in the experiment only one did so, makes for a slight overestimation of the effect for smooth window edges whilst it makes no difference when edges are sharp since the force only depends on the product of the reflection coefficients of the two materials. Note furthermore that the erratic behaviour of the integrand makes the numerical accuracy of integration somewhat rough, and also that in the modelling of the transparency band no effort has been made to ensure causality is satisfied. Thus the estimate is accurate only to order of magnitude.

The results for different values of $\Delta$ and $s$ at separation $a=100$nm are shown in figure \ref{fig_dP}. Note that the absolute pressure between parallel gold plates at $T=0$ is approximately 6 Pa (e.g.\ \cite{brevik06}) and that the difference between using Drude and plasma models is negligible here. Clearly the corrections at $\Delta$ close to $1$ are much too large, more than 10 times the force itself. The smoothness of the edges has no obvious effect. Only the plasma model with sharp boundaries and $\Delta<0.5$ could fall within the experimental accuracy of \cite{iannuzzi04}, yet it seems highly likely that the realitvely small corrections at low $\Delta$ and high $s$ for the plasma model is due to chance, especially since the more realitic Drude model gives corrections which are enormous and also, counterintuitively, positive. Such large corrections would be in keeping with Ford's predictions, but seem clearly ruled out by experiment.

\begin{figure}[t]
  \centerline{\includegraphics[width=2.3in]{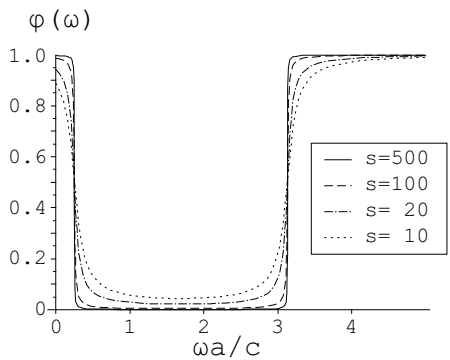}}
  \caption{$\varphi(\omega)$ for $\Delta=1$ with different $s$ values.}
  \label{fig_phi}
\end{figure}

\begin{figure}[t]
  \centerline{\includegraphics[width=2.5in]{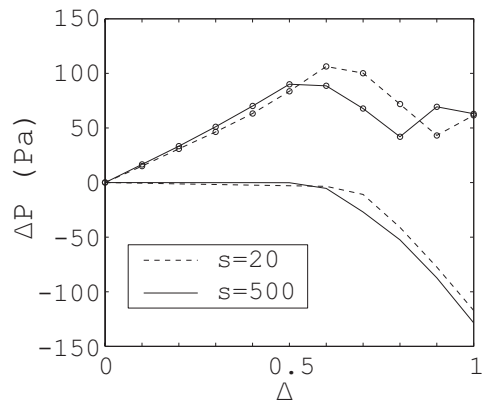}}
  \caption{Force difference with and without a transparent window using direct frequency calculation. Circles: Drude model, no circles: plasma model.}
  \label{fig_dP}
\end{figure}

\section{Conclusions}

We have revisited the question of the frequency spectrum of the Casimir force, generalising Ford's result \cite{ford93} to the case of subunitary reflection coefficients. While this smooths out the frequency spectrum as given by the real frequency Lifshitz formula integrand, the integrand is still wildly oscillating. This result is used to calculate the Casimir force and free energy for constant, ``effective'', reflection coefficients, a new result to the author's knowledge.

While the effective reflection coefficient may be a useful model, there is little reason to believe the resulting frequency spectrum to represent physics. A more realistic spectrum using physical models for the permittivity of materials is still highly irregular and, taken at face value, indicates that large and even repulsive Casimir forces could be attainable by tuning the dielectric response of materials used. An alternative means of calculation, paradoxically, gives a different, pessimistic result, and the large effects seemingly implied by the wildly behaved frequency spectrum will seem to be excluded by a recent experiment by Ianuzzi et alia. 

With the possibility of technological applications of the Casimir force however, there is reason to strive for a better understanding of the physical interpretation of the frequency spectrum of the Casimir force as well as further experimental efforts to settle this issue. An experiment similar to \cite{iannuzzi04} accompanied by a careful measurement of the dielectric response of the actual sample used over a large frequency region would be a straightforward possibility, and a more sensitive measurement of the force might also be able to measure the actual difference in pressure. 

It appears that the straightforward interpretation of the Casimir frequency spectrum as the integrand of the Lifshitz force formula at real frequencies is not valid, yet given the modesty of the efforts presented herein further investigation is warranted. Furthermore, the paradox presented herein pends a satisfactory resolution, hopefully to appear in the future.

The author thanks Irina Pirozhenko for supplying the data from \cite{palik95} in electronic format and acknowledges highly useful suggestions from an anonymous referee.

\end{document}